\documentclass[aps,prb,superscriptaddress,twocolumn,showpacs,amsmath,amssymb]{revtex4}
\usepackage{graphicx}
\usepackage{bm}

\begin{document}


\title{STM/STS study on large pseudogap and nodal superconducting gap in Bi2201(La) and Bi2212}


\author{T.~Kurosawa}
\affiliation{Department of Physics, Hokkaido University, Sapporo
060-0810, Japan}

\author{T.~Yoneyama}
\affiliation{Department of Physics, Hokkaido University, Sapporo
060-0810, Japan}

\author{Y.~Takano}
\affiliation{Department of Physics, Hokkaido University, Sapporo
060-0810, Japan}

\author{M.~Hagiwara}
\affiliation{Department of Physics, Hokkaido University, Sapporo
060-0810, Japan}

\author{R.~Inoue}
\affiliation{Department of Physics, Hokkaido University, Sapporo
060-0810, Japan}

\author{N.~Hagiwara}
\affiliation{Department of Physics, Hokkaido University, Sapporo
060-0810, Japan}

\author{K.~Kurusu}
\affiliation{Department of Physics, Hokkaido University, Sapporo
060-0810, Japan}

\author{K.~Takeyama}
\affiliation{Department of Physics, Hokkaido University, Sapporo
060-0810, Japan}

\author{N.~Momono}%
\affiliation{Department of Materials Science and Engineering,
Muroran Institute of Technology, Muroran 050-8585, Japan}

\author{M.~Oda}
\affiliation{Department of Physics, Hokkaido University, Sapporo
060-0810, Japan}

\author{M.~Ido}
\affiliation{Department of Physics, Hokkaido University, Sapporo
060-0810, Japan}


\date{\today}

\begin{abstract}
In the present work, scanning tunneling microscopy/spectroscopy (STM/STS) measurements were carried out on underdoped $\rm Bi_2Sr_{2-{\it x}}La_{\it x}CuO_{6+\delta}$ and $\rm Bi_2Sr_2CaCu_2O_{8+\delta }$ to clarify the origin of the pseudogap, in particular, the inhomogeneous large pseudogap. The nodal part of a $d$-wave pairing gap, which is under no influence of the inhomogeneous large pseudogap, was also examined by relating the homogeneous bottom part of the STS gap to a nodal $d$-wave gap in momentum space. We report that the inhomogeneous large pseudogap in the antinodal region links to a two-dimensional electronic charge order, and that the gap size of the nodal $d$-wave part $\rm {\Delta}_{sc}$ scales with the superconducting critical temperature $\rm {\it T}_c$ in the pseudogap regime.

\end{abstract}

\pacs{74.25.Jb, 74.50.+r, 74.72.Hs}

\maketitle

\section{Introduction}

To clarify the relationship between the pseudogap and the superconductivity, particularly whether the pseudogap is associated with incoherent precursor pairing of the superconductivity or an ordered state competing with the superconductivity, has been one of the central issues in the research field of high-$\rm {\it T}_c$ cuprate superconductors. The former case leads us to look at the antinodal region where the pseudogap phenomena are prominent as a source of attractive force causing the superconductivity, while the latter can provide us with a clue to understanding the long-standing puzzling problem of why $\rm {\it T}_c$ is largely reduced in the underdoped region. Earlier scanning tunneling microscopy/spectroscopy (STM/STS) and angle-resolved photoemission spectroscopy (ARPES) measurements carried out on slightly underdoped $\rm Bi_2Sr_2CaCu_2O_{8+\delta }$ (Bi2212) provided evidence for the possibility that the pseudogap resulted from incoherent precursor pairing.\cite{Loser,Renner,Miyakawa,Campuzano}

On the other hand, Vershinin {\it et al.} reported from STM/STS studies on Bi2212 that a two-dimensional (2D) electronic charge order, the so-called checkerboard charge order, appeared in the pseudogap state, and claimed that the 2D charge order was a possible hidden order of the pseudogap state.\cite{Vershinin} Hanaguri {\it et al.} also found a similar 2D charge order in STM studies on the pseudogap state of lightly doped $\rm Ca_{2-{\it x}}Na_{\it x}CuO_2Cl_2$.\cite{Hanaguri} Following these works, Kohsaka {\it et al.} reported that the spatial structure of the 2D charge order consisted of 4$a$-wide unidirectional domains ($a$: lattice constant) that oriented randomly along two equivalent Cu-O bonds without long-range order.\cite{Kohsaka2007s} Recently, ARPES studies on deeply underdoped Bi2212, underdoped $\rm Bi_2Sr_{2-{\it x}}La_{\it x}CuO_{6+\delta }$ (Bi2201(La)) and $\rm La_{2-{\it x}}Sr_{\it x}CuO_4$ (La214) have demonstrated that the pseudogap survives down to below $\rm {\it T}_c$ with a gap size $\rm {\Delta}^*$ much larger than that of the $d$-wave pairing gap ${\Delta}_0$,\cite{Tanaka,Kondo2007prl,Terashima2005prl} indicating that such a large pseudogap of the order $\rm {\Delta}^*$ $(\textgreater {\Delta}_0$) is independent of incoherent precursor pairing. If the large pseudogap of the order $\rm {\Delta}^*$ results from the 2D charge order, the ARPES observation is consistent with STM/STS observations that the same 2D charge order appears below and above $\rm {\it T}_c$ in Bi2212\cite{Vershinin,Howald,Liu} and Bi2201(La)\cite{Machida,Saito,Wise}. Furthermore, recent STS and ARPES studies on Bi2212 and Bi2201(La) have revealed that a nodal part of the $d$-wave gap starts to open below $\rm {\it T}_c$ while the gap size of the large pseudogap shows no clear change across $\rm {\it T}_c$, implying that the nodal $d$-wave part links to the superconductivity directly.\cite{Lee2007,Boyer,Kondo2009,Pushp}

Recently, ARPES experiments on nearly optimally doped high-$\rm {\it T}_c$ cuprates such as Bi2201(La) and La214 found two different types of pseudogaps in the antinodal region, a large pseudogap and a so-called small pseudogap whose gap size is comparable to that of the $d$-wave gap ${\Delta}_0$.\cite{Kondo2007prl,Terashima2005prl,Wei,Yoshida2007,Shi} The small pseudogap and the nodal $d$-wave gap evolving below $\rm {\it T}_c$ are integrated into a single $d$-wave gap at $\rm {\it T} $$\ll $${\it T}_c$, suggesting that the small pseudogap results from incoherent precursor pairing or that the small pseudogap is closely related to the superconducting (SC) gap. However, the reason why two pseudogaps with different energy scales appear exclusively in the antinodal region is open to question. 

In the present work, to clarify the origin of the pseudogap, in particular the large pseudogap, we performed STM/STS measurements on underdoped Bi2201(La) and Bi2212 at $\rm {\it T} $$\ll $${\it T}_c$, and confirmed that the large pseudogap inhomogeneously spatially links to the static 2D electronic charge order. We also report that the nodal part of the $d$-wave gap, which is free from the large pseudogap, is associated with the homogeneous bottom part of the STS gap, and its gap size $\rm {\Delta}_{sc}$ correlates with the SC critical temperature $\rm {\it T}_c$. 

\section{Experimental Procedures}

Both Bi2201(La) and Bi2212 crystals were grown by using the traveling solvent floating zone method. We controlled doping level $p$ by changing the pressure of the oxygen atmosphere in the course of growing the crystals. In preparation of Bi2201(La) crystals, we substituted La$^{3+}$ for a part of Sr$^{2+}$ in order to reduce doping level $p$. In the present work, the SC critical temperature $\rm {\it T}_c$ was defined by extrapolating the steepest part of the SC diamagnetic curve to the zero level. We performed low-bias STM imaging, which enabled us to observe the Cu-O layer buried below the cleaved Bi-O layer, on Bi2201(La) and Bi2212 samples at 8$\sim $9 K. We cleaved crystals in situ in an ultra high vacuum just before moving the STM tip toward the cleaved surface of the sample. Details of the low-bias STM imaging and STS measurements have been reported elsewhere.\cite{Momono2005j,Hashimoto}

\section{Results and Discussion}

\subsection{STM images of Bi2201(La) and Bi2212}

\begin{figure}[t]
\begin{center}
\includegraphics{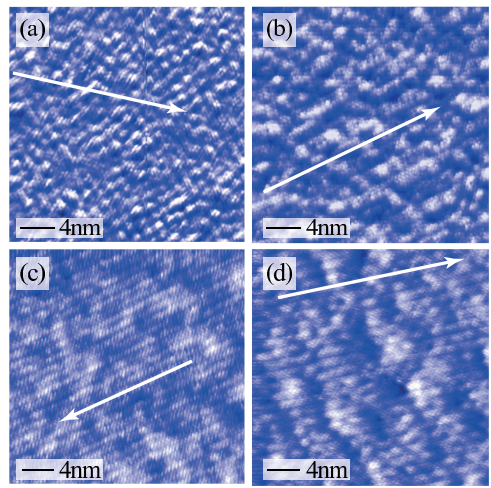}
\end{center}
\caption{(Color online) Low-bias STM images of (a) Bi2201(La) sample A, (b) Bi2201(La) sample B and (c) Bi2201(La) sample C, measured at a bias voltage of $\rm {\it V}_b$=20 mV and a tunneling current of $\rm {\it I}_t$=0.08 nA at $T $$\sim$9 K. (d) Low-bias STM image of Bi2212 sample A, measured at $\rm {\it V}_b$=30 mV and $\rm {\it I}_t$=0.07 nA at $T $$\sim$8 K.}
\label{STM}
\end{figure}

\begin{figure}[t]
\begin{center}
\includegraphics{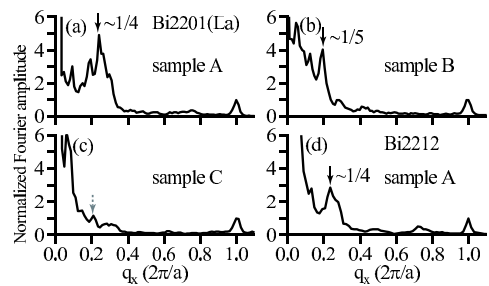}
\end{center}
\caption{Line cuts of 2D Fourier maps of the STM images along the (0, 0)-($\pi$, 0) direction for (a) Bi2201(La) sample A, (b) Bi2201(La) sample B, (c) Bi2201(La) sample C and (d) Bi2212 sample A. Here, the amplitude of the line profile is normalized by the height of the Bragg peak. Arrows show the 2D charge order's Fourier peak for each sample.}
\label{FT}
\end{figure}

Figures\ \ref{STM}(a), (b) and (c) show low-bias STM images taken on underdoped Bi2201(La) sample A ($x$=0.6, $\rm {\it T}_c$=25 K, $p $$\sim$0.10), and nearly optimally doped Bi2201(La) sample B ($x$=0.4, $\rm {\it T}_c$=32 K, $p $$\sim$0.14) and C ($x$=0.4, $\rm {\it T}_c$=32 K, $p $$\sim$0.14) at a bias voltage of $\rm {\it V}_b$=20 mV and a tunneling current of $\rm {\it I}_t$=0.08 nA. A typical low-bias STM image is also shown for underdoped Bi2212 sample A ($\rm {\it T}_c$=64 K, $p $$\sim$0.10) in Fig.\ \ref{STM}(d). A bond-oriented 2D charge order clearly appears throughout the STM images of Bi2201(La) samples A and B as well as in Bi2212 sample A. The 2D charge order is nondispersive in the sense that its period ${\it \lambda}$ is independent of bias voltage $\rm {\it V}_b$, and $\it \lambda$ depends on the doping level $p$, at least, in Bi2201(La); $\it \lambda $$\sim$4$a$ in Bi2201(La) sample A as well as Bi2212 sample A, but $\sim$5$a$ in Bi2201(La) sample B (Fig.\ \ref{FT}). Such a change of ${\it \lambda}$ is consistent with the $p$-dependence of $\it \lambda$ reported by Wise {\it et al.}\cite{Wise}

On the other hand, low-bias STM images of Bi2201(La) sample C, which were obtained from the same Bi2201(La) rod as sample B, exhibit no clear 2D charge order, as seen in Figs.\ \ref{STM}(c) and\ \ref{FT}(c). Such a strong sample (or cleaved surface) dependence of the charge-order amplitude implies that the 2D charge order will be highly sensitive to disorder probably introduced by lattice imperfections or dopant atoms. The strong sample (or cleaved surface) dependence of the 2D charge order is true of underdoped Bi2212, and has been discussed in terms of pinning of a dynamical 2D electronic charge order from the point of view that the 2D electronic charge order will be dynamical in itself.\cite{Momono2005j,Hashimoto} 

\subsection{STS spectra of Bi2201(La) and Bi2212}

\begin{figure}[t]
\begin{center}
\includegraphics[scale=.60]{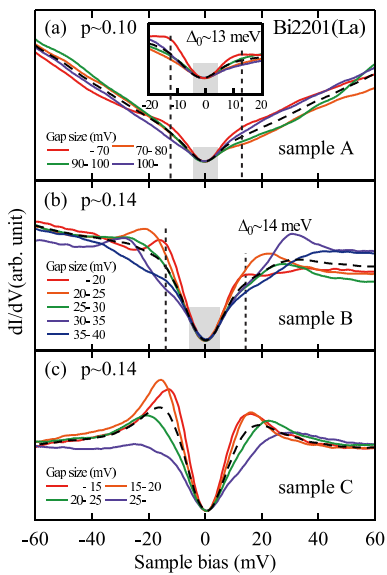}
\end{center}
\caption{(Color) STS spectra for (a) Bi2201(La) sample A ($x$=0.6, $\rm {\it T}_c$=25 K, $p $$\sim$0.10), (b) sample B ($x$=0.4, $\rm {\it T}_c$=32 K, $p $$\sim$0.14) and (c) sample C ($x$=0.4, $\rm {\it T}_c$=32 K, $p $$\sim$0.14), taken along the white lines in Figs.\ \ref{STM}(a)-(c). STS spectra were classified into several groups with different gap widths and the classified spectra were averaged among each group. The dashed line shows the averaged spectrum, obtained by averaging all STS spectra of each sample. The shaded area covers the homogeneous bottom part of the STS spectrum. The inset of (a) shows STS spectra, measured from other parts of sample A.}
\label{STS01}
\end{figure}

\begin{figure}[t]
\begin{center}
\includegraphics[scale=.60]{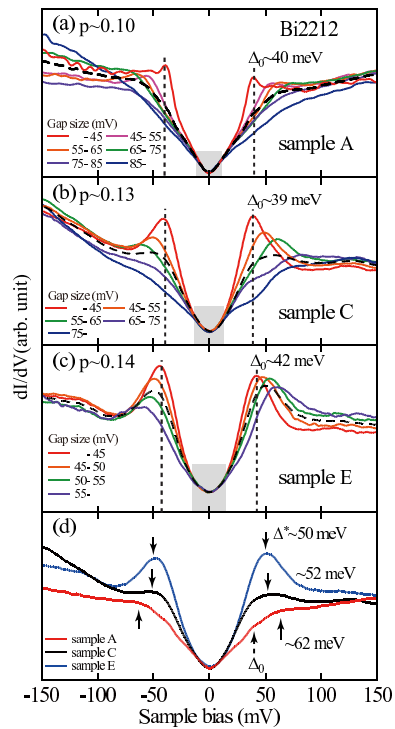}
\end{center}
\caption{(Color) Classified and averaged STS spectra with different gap widths for (a) Bi2212 sample A ($\rm {\it T}_c$=64 K, $p $$\sim$0.10), taken along the white line in Fig.\ \ref{STM}(d). Classified and averaged STS spectra of (b) Bi2212 samples C ($\rm {\it T}_c$=78 K, $p $$\sim$0.13) and (c) E ($\rm {\it T}_c$=81 K, $p $$\sim$0.14), whose original spectra before averaging were previously reported in ref. (24). The shaded area covers the homogeneous bottom part of the STS spectrum. (d) The averaged spectrum obtained by averaging all STS spectra of each sample. The STS spectrum thus obtained is also shown by dashed lines in Figs.\ \ref{STS12}(a)-(c).}
\label{STS12}
\end{figure}

STS spectra ($dI/dV$ curves) of Bi2201(La) samples A and B, and Bi2212 sample A, obtained at 8$\sim$9 K along the lines marked on the STM images (Fig.\ \ref{STM}), are very inhomogeneous spatially. Therefore the STS spectra of each sample were classified into several groups with different gap widths, and the classified spectra were averaged among each group, as shown in Figs.\ \ref{STS01}(a)-(c) and\ \ref{STS12}(a). Classified and averaged STS spectra of Bi2212 samples C ($\rm {\it T}_c$=78 K, $p $$\sim$0.13) and E ($\rm {\it T}_c$=81 K, $p $$\sim$0.14), whose original STS spectra before averaging were reported in ref. (24), are also shown in Figs.\ \ref{STS12}(b) and (c). In Bi2201(La) samples A and B, and Bi2212 sample A, STM images of samples that exhibit strong 2D charge orders, features of the gap structure such as gap width and gap edge peak height largely vary with the STS measurement position. Furthermore, a sub gap structure appears inside the gap edge peaks of the STS gap whose width is relatively broad. However, it is noteworthy that the bottom part of the STS gap (the shaded range of $\rm {\it V}_b$ in Figs.\ \ref{STS01}(a), (b) and\ \ref{STS12}(a)-(c)) is homogeneous although the overall STS gap is rather inhomogeneous, as was already reported in previous STS studies on underdoped Bi2212.\cite{Hashimoto,McElroy2005,Liu2008} This means that quasiparticle states associated with the nodal parts of the $d$-wave gap, which dominate the bottom part of the STS gap, are homogeneous. Therefore, the inhomogeneity of the overall STS gap should be attributable to the nature of quasiparticles around the antinodal region, where the pseudogap develops. 

Another important observation is that the gap structure of STS spectra is rather different between Bi2201(La) sample B exhibiting a strong 2D charge order in its STM image and Bi2201(La) sample C exhibiting a very weak one, although both samples B and C were obtained from the same single-crystal rod. The great majority of STS spectra of sample C show a gap structure of the $d$-wave type with sharp gap edge peaks and no sub gap. This is in sharp contrast to STS data of Bi2201(La) sample B, whose STS gap is characterized by inhomogeneous, broad gap width and the sub gap structure, as mentioned above. Such a contrasting feature of the STS gap between these samples is attributable to the different natures of their static 2D charge orders. We will focus on this fact again in next subsection in order to discuss the origin of the large pseudogap. 

\subsection{Relationship between the large pseudogap and the static 2D charge order}

\begin{figure}[h]
\begin{center}
\includegraphics[scale=.6]{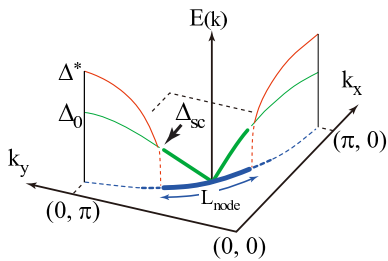}
\end{center}
\caption{(Color online) Schematic illustration of two different types of gap on the Fermi surface. One is the large pseudogap, whose gap width $\rm \Delta^*$ is much larger than that of $d$-wave gap, and the other is the small pseudogap, whose gap width $\rm \Delta_0 $ is comparable to that of simple $d$-wave gap. Both gaps evolve on the antinodal parts of Fermi surface in momentum space.}
\label{twogap}
\end{figure}

Recently, ARPES experiments on optimally doped Bi2201(La) demonstrated that two different gap structures exist, with a single component and two components respectively, at $\rm {\it T} $$\ll $${\it T}_c$.\cite{Wei} The two-component gap, which was first found in ARPES measurements on lightly doped Bi2212 and nearly optimally doped Bi2201(La), consists of a $d$-wave gap formed on the nodal part of the Fermi surface and the large pseudogap formed on the antinodal one whose gap width $\rm \Delta^* $ is much larger than that of the $d$-wave gap.\cite{Tanaka,Kondo2007prl} On the other hand, the single-component gap follows a $d$-wave gap function over the entire Fermi surface. The antinodal part of the single-component gap, whose gap width $\rm \Delta_0 $ is comparable to that of the $d$-wave gap appearing in the two-component gap, smoothly changes into the small pseudogap at $\rm {\it T} \textgreater {\it T}_c$, as schematically shown in Fig.\ \ref{twogap}.

\begin{figure}[b]
\begin{center}
\includegraphics[scale=.6]{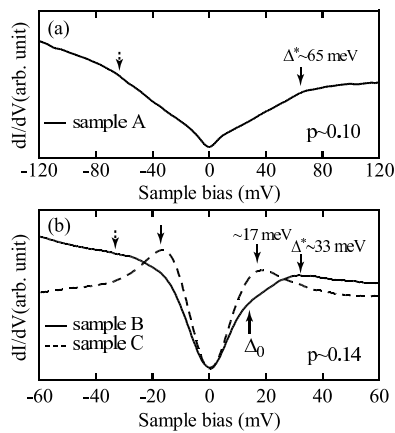}
\end{center}
\caption{(a), (b) The averaged spectrum obtained by averaging all STS spectra of Bi2201(La) samples A, B and C are also shown by dashed lines in Figs.\ \ref{STS01}(a)-(c).}
\label{STS01ave}
\end{figure}

\begin{figure}[t]
\begin{center}
\includegraphics[scale=.5]{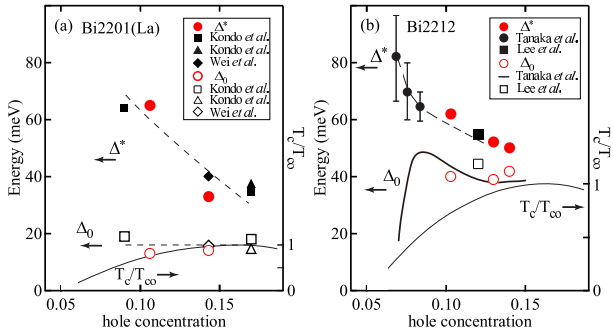}
\end{center}
\caption{(Color) (a) Doping-level dependence of $\rm \Delta^*$ (closed red circles) and $\rm \Delta_0$ (open red circles) of Bi2201(La), plotted together with ARPES data $\rm {\Delta}_{ARPES}^*$ and $\rm {\Delta}_0^{ARPES}$ reported by Kondo {\it et al.} (closed and open triangles)\cite{Kondo2007prl}, Kondo {\it et al.} (closed and open squares)\cite{Kondo2009} and Wei {\it et al.} (closed and open rhombi)\cite{Wei}. (b) Doping-level dependence of $\rm \Delta^*$ (closed red circles) and $\rm \Delta_0$ (open red circles) of Bi2212, plotted together with ARPES data $\rm {\Delta}_{ARPES}^*$ and $\rm {\Delta}_0^{ARPES}$ reported by Tanaka {\it et al.} (closed black circles and solid line)\cite{Tanaka} and Lee {\it et al.} (closed and open squares)\cite{Lee2007}. The scale for $\rm {\it T}_c$ (the right-hand axis) is normalized by the optimal value of the critical temperature $\rm {\it T}_{co}$.}
\label{delta}
\end{figure}

To compare the present STS data with ARPES data, which reflect the electronic structure averaged over the entire cleaved surface, we averaged all STS spectra of each sample (Figs.\ \ref{STS01} and\ \ref{STS12}). Figures\ \ref{STS12}(d) and\ \ref{STS01ave} show STS spectra thus averaged for both Bi2212 and Bi2201(La) samples. The averaged STS spectra for Bi2201(La) samples A ($p $$\sim$0.10) and B ($p $$\sim$0.14), exhibiting strong 2D charge orders in their STM images, show broad peaks at energies (bias voltages) of $\rm \Delta^* $$\sim$65 meV and $\sim$33 meV, respectively, although the peak structure is not so clear on the negative bias side, especially in sample B. It is noteworthy here that these peak energies are comparable to the gap width of the large pseudogap $\rm \Delta^*_{ARPES}$ reported by ARPES studies on underdoped Bi2201(La) samples with $p $$\sim$0.09 ($\rm \Delta^*_{ARPES} $$\sim$64 meV)\cite{Kondo2009} and nearly optimally doped Bi2201(La) samples with $p $$\sim$0.14 ($\rm \Delta^*_{ARPES} $$\sim$40 meV)\cite{Wei}, as shown in Fig.\ \ref{delta}(a). Such agreement between $\rm \Delta^*$, estimated from half the distance between the broad peak of the averaged STS spectra, and $\rm \Delta^*_{ARPES}$ is also true of underdoped Bi2212 samples A, C and E, whose STM images also exhibit strong 2D charge orders (Fig.\ \ref{delta}(b)).\cite{Tanaka,Lee2007} The agreement of $\rm \Delta^*$ and $\rm \Delta^*_{ARPES}$ in both Bi2201(La) and Bi2212 systems means that the broad peak of the averaged STS gap corresponds to the large pseudogap. 

\begin{figure}[h]
\begin{center}
\includegraphics[scale=.6]{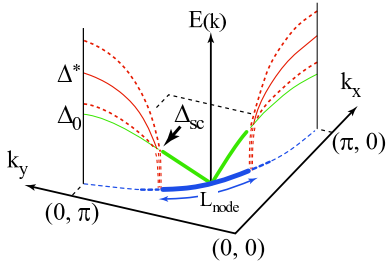}
\end{center}
\caption{(Color) Schematic illustration of the gap structure over the entire Fermi surface. The green line represents a $d$-wave gap. Orange dotted lines represent a spatially inhomogeneous large pseudogap, and orange solid lines the pseudogap averaged over the entire cleaved surface. Corresponding to the observation that inhomogeneous dispersion curves of the large pseudogap tend to converge at a certain bias voltage (energy) in each sample (Figs.\ \ref{STS01}(a), (b) and\ \ref{STS12}(a)-(c)), the large pseudogap is emphatically drawn to evolve within a definite region of momentum space regardless of the large pseudogap size.}
\label{FS}
\end{figure}

On the other hand, the averaged STS gap of Bi2201(La) sample C (Fig.\ \ref{STS01ave}(b)) shows no large pseudogap; its peak energy (17$\sim $18 meV) is comparable to the gap width of the single-component gap $\rm \Delta^{ARPES}_0$ (=14$\sim $18 meV) reported by ARPES measurements on optimally and nearly optimally doped Bi2201(La) samples,\cite{Kondo2007prl,Kondo2009,Wei} as shown in Fig.\ \ref{delta}(a). STM images of sample C exhibit very weak 2D charge orders whereas those of sample B, obtained from the same single-crystal rod as sample C, have strong 2D charge orders, as mentioned above. Therefore, the lack of the appearance of the large pseudogap in sample C provides us with evidence that the large pseudogap is linked to the static 2D charge order directly. (Bi2212 samples, whose STM images exhibit very weak 2D charge orders, also show no large pseudogaps, as seen in Figs. 13 and 16 of ref. 24.) Furthermore, in STS data set shown in Figs.\ \ref{STS01} and\ \ref{STS12} we notice that the gap width of the large pseudogap $\rm \Delta^*$ largely varies with the measurement position of STS, though the large pseudogap appears above a certain bias voltage regardless of $\rm \Delta^*$ within a sample, indicating that it will evolve over a definite region in momentum space. Those features of the large pseudogap, schematically shown in Fig.\ \ref{FS}, are in consistent with recent STS data on Bi2212 reported by Pushp {\it et al}.,\cite{Pushp} and could be attributable to the short-range static 2D charge order. 

\begin{figure}[t]
\begin{center}
\includegraphics[scale=.7]{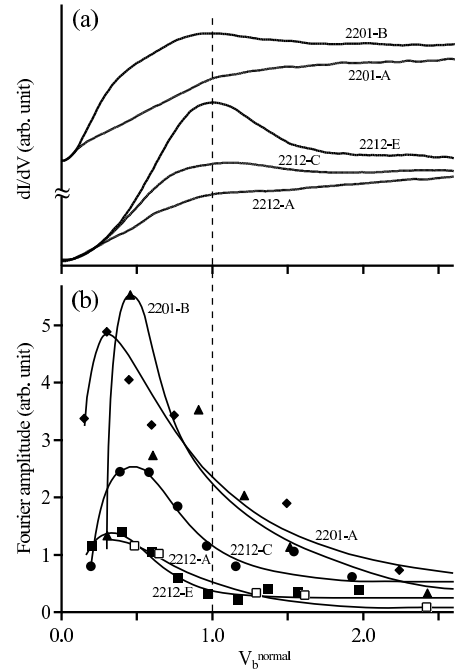}
\end{center}
\caption{(a) Averaged STS spectra of Bi2201(La) samples A and B, and Bi2212 samples A, C, and E, shown in Figs.\ \ref{STS12}(d) and\ \ref{STS01ave}, are plotted on the positive side of bias voltage ($\rm {\it V}_b \textgreater $0). (b) Energy (bias voltage $\rm {\it V}_b$) dependence of the Fourier peak intensity of the 2D charge order is plotted as a function of $\rm {\it V}_b$ for Bi2201(La) samples A and B, and Bi2212 samples A, C, and E. Both horizontal scales in (a) and (b) are normalized by the large pseudogap width $\rm \Delta^*$(=e$\rm {\it V}_b^*$) for each sample; $\rm {\it V}_b^{normal}$ =$\rm {\it V}_b$/$\rm {\it V}_b^*$.}
\label{intens}
\end{figure}

The interrelation between the large pseudogap and the static 2D charge order can also be confirmed in the following experimental result. In Figs.\ \ref{intens}(a) and (b), the averaged gap structures (Figs.\ \ref{STS12}(d) and\ \ref{STS01ave}) and the peak intensity of the Fourier spot of the static 2D charge order are shown for both Bi2201(La) and Bi2212 samples as a function of bias voltage $\rm {\it V}_b$ on the positive bias side. (The horizontal scales in Figs.\ \ref{intens}(a) and (b) are normalized by the large pseudogap width $\rm \Delta^*$(=e$\rm {\it V}_b^*$) for each sample; $\rm {\it V}_b^{normal}$ =$\rm {\it V}_b$/$\rm {\it V}_b^*$.) We note in Figs.\ \ref{intens}(a) and (b) that the 2D charge order appears at bias voltages lower than $\rm \Delta^*$/e, namely within the large pseudogap, but it is markedly suppressed at very low bias voltages corresponding to the homogeneous nodal part of the $d$-wave gap. Such a result also implies that the large pseudogap is intimately related to the static 2D charge order. 

A similar relationship was already reported by McElroy {\it et~al.} for the static 2D charge order locally appearing at very high bias voltages above $\rm {\it V}_b$=65 mV and a very large pseudogap larger than 65 meV (the so-called zero-temperature pseudogap with no peak structure: ZTPG) in underdoped Bi2212.\cite{McElroy2005} For $\rm {\it V}_b \textless 65$ mV, they reported a dispersive 2D charge order (spatial structure of the electronic density) resulting from the scattering interference effect of SC-quasiparticles, but no nondispersive 2D charge order.\cite{McElroy2005,Hoffman,McElroy2003}

\subsection{Sub gap anomaly inside the large pseudogap}

In Bi2212 and Bi2201(La) samples exhibiting strong 2D charge orders in their STM images, the sub gap structure appears as a plateau or a shoulder inside the large pseudogap, although it is difficult to identify the sub gap structure on the negative bias side (Figs.\ \ref{STS01}(a), (b) and\ \ref{STS12}(a)-(c)). We note that the sub gap position is roughly in agreement with the peak position of the narrowest gap with no sub gap on the positive bias side, as seen in Figs.\ \ref{STS01}(a), (b) and\ \ref{STS12}(a)-(c). Furthermore, the energy of the sub gap position on the positive bias side or half the distance between the narrowest gap peaks of Bi2212, referred to as $\rm {\Delta}_0$ in Fig.\ \ref{STS12}, corresponds to the antinodal $d$-wave gap $\rm {\Delta}_0^{ARPES}$, which was obtained by extrapolating the nodal part of the $d$-wave gap to the antinodal point in ARPES measurements on Bi2212, as shown in Fig.\ \ref{delta}(b).\cite{Tanaka,Lee2007} In Bi2201(La), $\rm {\Delta}_0$ and $\rm {\Delta}_0^{ARPES}$ also roughly agree with each other (Fig.\ \ref{delta}(a)).\cite{Kondo2007prl,Kondo2009,Wei}

The features of the sub gap structure mentioned above lead us to the idea that, if the large pseudogap evolves in the antinodal region at high temperatures above $\rm {\it T}_c$, the density of states inside the large pseudogap is reduced to a large degree but still remains finite. This finite in-gap density of states $\rm {\it N}_{anti}(0)$ will allow another gap (a sub gap) to open inside the large pseudogap at a lower temperature $\rm {\it T}_c$. This idea, which is the same as the ``soft gap'' proposed by Ma {\it et al.} for the pseudogap,\cite{Ma} can explain the spatial change between the sub gap structure and the narrowest gap with a width of the order $\rm {\Delta}_0$ as follows: the sub gap structure turns into the narrowest gap over the region where the large pseudogap happens to develop insufficiently and suppresses $\rm {\it N}_{anti}(0)$ only slightly. On the other hand, the narrowest gap is reduced to a sub gap structure over the region where the large pseudogap develops moderately and suppresses $\rm {\it N}_{anti}(0)$ to some extent. Such a scenario is consistent with the recent ARPES observation on Bi2201(La) that the losses of spectral weight arising from the pseudogap and the $d$-wave gap evolution compete with each other over the antinodal region in momentum space.\cite{Kondo2009}

\subsection{Nodal superconducting gap}

In simple $d$-wave superconductors within the Bardeen-Cooper-Schrieffer regime, the antinodal part of the $d$-wave gap $\rm {\Delta}_0$, reflecting pairing strength, scales with $\rm {\it T}_c$. However, in underdoped Bi2201 and Bi2212 samples, in which the weight of the antinodal part of the $d$-wave gap is largely suppressed through the competition with the large pseudogap, $\rm {\Delta}_0$ does not scale with $\rm {\it T}_c$ (Fig.\ \ref{delta}). As was demonstrated in ARPES measurements on nearly optimally doped Bi2212, the antinodal part of the $d$-wave gap, which will smoothly change into the small pseudogap at $\rm {\it T} \textgreater {\it T}_c$, shows no typical signature of the coherent SC state such as Bogoliubov quasi-particles even at $\rm {\it T} $$\ll $${\it T}_c$.\cite{Lee2007} Thus, besides the small weight of the antinodal $d$-wave part, the incoherent nature of the antinodal $d$-wave part may be related to the breakdown of the scaling between $\rm {\Delta}_0$ and $\rm {\it T}_c$. Therefore we examined the relationship between the nodal $d$-wave part and $\rm {\it T}_c$. 

\begin{figure}[h]
\begin{center}
\includegraphics[scale=.70]{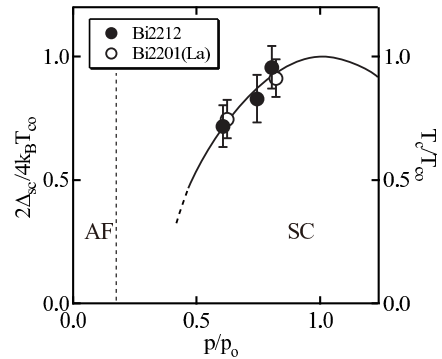}
\end{center}
\caption{The gap sizes of the coherent nodal part $\rm {\Delta}_{sc}$ for Bi2212 (closed circles) and Bi2201(La) (open circles). Here, the vertical axis and horizontal axis are normalized by the optimal value of the critical temperature $\rm {\it T}_{co}$ and the optimal doping level $\rm {\it p}_o$ respectively.}
\label{Tc}
\end{figure}

In samples exhibiting the large pseudogap, the STS gap is homogeneous at the bottom, whereas it is very inhomogeneous outside the bottom part on account of the large pseudogap which is sensitive to disorder (Fig.\ \ref{FS}), as mentioned above. This allows us to estimate the $d$-wave gap size at the edge of the nodal Fermi surface, $\rm {\Delta}_{sc}$, from half the width of the homogeneous bottom part (the shaded range of $\rm {\it V}_b$) in Figs.\ \ref{STS01} and\ \ref{STS12}. In Fig.\ \ref{Tc}, $\rm {\Delta}_{sc}$ thus obtained and $\rm {\it T}_c$ are plotted with the same ratio for Bi2212 and Bi2201(La): 2$\rm {\Delta}_{sc}$/$\rm {\it k}_B{\it T}_c$=4. A noteworthy fact here is that $\rm {\Delta}_{sc}$ scales with $\rm {\it T}_c$ in the underdoped regions of both systems, which is in agreement with recent ARPES results reported by Yoshida {\it et al.} for La214 and Bi2212.\cite{Yoshida2009} Such scaling supports scenarios of nodal superconductivity. Presumably, only mobile carriers (holes) on the nodal Fermi surface, which have high in-plane mobility and dominate the transport properties of the Cu-O plane,\cite{Yanase} will play a crucial role in causing the superconductivity.\cite{Pines,Lee,Geshkenbein,Wen,Furukawa,Ido1998,Oda2000} 

\subsection{Superconducting energy scale determining $\rm {\it T}_c$ in bulk}

Finally, we pay attention to the high sensitivity of the static 2D charge order and/or the large pseudogap to disorder, because its nature leads to the possibility that the static (pinned) 2D charge order and the large pseudogap are surface phenomena in the case of clean samples. In that case, the bulk gap structure will be of a single $d$-wave type with no large pseudogap at $\rm {\it T} $$\ll $${\it T}_c$, and its antinodal part, which might be associated with the dynamical 2D electronic charge order, will survive as a small pseudogap at $\rm {\it T} \textgreater {\it T}_c$. In the present STS study, we could virtually identify the narrowest gap with that of the bulk, because the narrowest gap accompanied no large pseudogap or a very weak one if any. However, the antinodal gap size $\rm {\Delta}_0$ ($\sim \rm {\Delta}_0^{bulk}$) of the narrowest gap showed no scaling with $\rm {\it T}_c$ in the underdoped region, as mentioned above (Fig.\ \ref{delta}). Presumably, the nodal $d$-wave part will determine $\rm {\it T}_c$, independently of the antinodal part, even in bulk.

\section{Summary}

The present STM/STS results on Bi2201(La) and Bi2212 show that the inhomogeneous large pseudogap in the antinodal region originates in the static (pinned) 2D electronic charge order, consistent with the report by McElroy {\it et~al.} from STM/STS studies on Bi2212.\cite{McElroy2005} The large pseudogap and the static 2D charge order are highly sensitive to disorder, which provides a natural explanation for the quite contrasting observations of the large pseudogap in recent ARPES experiments; some ARPES experiments reported the observation of a large pseudogap but some others did not observe it.\cite{Kondo2007prl,Terashima2005prl,Wei,Yoshida2007,Shi}  In the present work, it was also pointed out that the gap size $\rm {\Delta}_{sc}$ of the nodal $d$-wave part, which is free from the large pseudogap or the small pseudogap, provides the SC energy scale determining $\rm {\it T}_c$. This is consistent with electronic Raman scattering experiments on high-$\rm {\it T}_c$ cuprates.\cite{Tacon} The SC energy scale $\rm {\Delta}_{sc}$ can also explain the marked suppression of the superconducting condensation energy in the pseudogap regime, as previously reported.\cite{Momono2002,Matsuzaki}

\vspace{1\baselineskip}
\begin{center}
{\bf Acknowledgments}
\end{center}
\vspace{1\baselineskip}

Thanks to useful discussions for F. J. Ohkawa. This work was supported in part by a Grant-in-Aid for Scientific Research and the 21st century COE program ``Topological Science and Technology'' from the Ministry of Education, Culture, Sports, Science, and Technology of Japan.

\end{document}